# Discrete modes of social information processing predict individual behavior of fish in a group

Roy Harpaz[1], Gašper Tkačik[2], and Elad Schneidman[1]

[1]Department of Neurobiology, Weizmann Institute of Science, Rehovot, Israel
[2]Institute of Science and Technology, Klosterneuburg, Austria

Individual computations and social interactions underlying collective behavior in groups of animals are of great ethological, behavioral, and theoretical interest. While complex individual behaviors have successfully been parsed into small dictionaries of stereotyped behavioral modes, studies of collective behavior largely ignored these findings; instead, their focus was on inferring single, mode-independent social interaction rules that reproduced macroscopic and often qualitative features of group behavior. Here we bring these two approaches together to predict individual swimming patterns of adult zebrafish in a group. We show that fish alternate between an 'active' mode in which they are sensitive to the swimming patterns of conspecifics, and a 'passive' mode where they ignore them. Using a model that accounts for these two modes explicitly, we predict behaviors of individual fish with high accuracy, outperforming previous approaches that assumed a single continuous computation by individuals and simple metric or topological weighing of neighbors' behavior. At the group level, switching between active and passive modes is uncorrelated among fish, yet correlated directional swimming behavior still emerges. Our quantitative approach for studying complex, multi-modal individual behavior jointly with emergent group behavior is readily extensible to additional behavioral modes and their neural correlates, as well as to other species.

## Introduction

Group behavior has been studied in a wide range of species – bacteria *(1)*, slime mold *(2)*, insects *(3–5)*, fish *(6–13)*, birds *(14–16)*, and mammals *(17–20)* – seeking the design principles of collective information processing, decision making, and movement. Theoretical models have suggested possible classes of computations and interaction rules that generate complex collective behavior, qualitatively replicate macroscopic features of behavior observed in real animal groups *(21–25)*, and also have algorithmic, behavioral, and economic implications *(26)*. The ability to record the movement patterns of animals in a group with high temporal and spatial precision for long periods *(10, 14, 16, 17, 27, 28)*, allows for direct exploration of individual traits and interactions between group members. Such attempts have considered topological vs. metric relations between conspecifics *(16)*, effective social "forces" depending on the distance between individuals *(6, 7)*, inference of functional interactions based on maximum entropy models of observed directional correlations *(15),* hierarchical spatial ordering *(14, 29, 30)*, and active signaling *(3, 31, 32)*.

Because individual behavior is complex, previous studies have mostly focused on modeling various group level statistics, e.g., polarization or moments of the distribution of inter-individual distances *(6–8, 16, 33, 34)*; see also *(35))*. These approaches, however, do not necessarily yield a unique solution for the underlying interactions between individuals *(35)*. Furthermore, the resulting models were often non-physiological in terms of response times or temporal causality, ignored physical constraints such as momentum and friction, or omitted the role of non-social sensory information. Somewhat surprisingly, most models of individual behavior in a group commonly assume that animals continuously update their movement based on the location or velocity of their neighbors *(22, 24, 25)*. In contrast, characterization of movement patterns of individual zebrafish larva, *C. elegans,* and *Drosophila*, for example, suggest that a distinct and relatively small set of stereotyped modes underlies complex individual behavior *(36–38)*. Here, we ask how discrete behavioral modes at the level of the individual affect sensory and social information processing underlying group-level motion decisions.

We studied individual behavior in groups of adult zebrafish in a large arena, using high spatio-temporal tracking of fish under different behavioral contexts. The adult zebrafish live in nature in groups of 4-20 fish either in still waters or in running rivers *(39)*, exhibit social behaviors and shoaling tendencies both in the wild and in the laboratory *(9, 39)* (unlike the transparent larvae that allow for imaging neuronal circuits underlying sensory-motor processing *(40–43)* but exhibit a limited behavioral repertoire *(44))*. We analyze the behavior of individuals in the group and identify distinct behavioral modes, which are used to build a highly accurate mathematical model of swimming behavior of individual fish in a group. The model is based on the sensory and social information that is available to each animal and takes into account spatial and temporal biophysical constraints. Importantly, we evaluate the models in terms of their power to predict individual fish trajectories, rather than statistical averages over the whole group.

**Results**

To study individual computations and interactions underlying group behavior in zebrafish, we tracked individuals in groups of 2, 3, and 6 adult fish for up to an hour at a time, in a large circular arena with shallow waters constituting an effective 2D environment (Fig. 1A, and S1A, Movie M1, SI Methods). We sampled the trajectory of the center of mass of each fish $i$ in the group, denoted as $\vec{x}_i(t)$, with high spatial and temporal resolution (see SI Methods). Decomposing the time-dependent velocity of each fish, $\vec{v}_i(t)$, into instantaneous swimming speed, $s_i(t) = |\vec{v}_i(t)|$, and instantaneous direction $\vec{d}_i(t) = \frac{\vec{v}_i(t)}{|\vec{v}_i(t)|}$, revealed a clear segmentation of the trajectories into acceleration and deceleration epochs (Fig. 1B-C, Movie M2). Acceleration epochs of the fish were very accurately described by a family of sigmoid functions that differed by their slope and duration (Fig. 1C-D). Decelerations were very accurately described by a single exponential, corresponding to a simple drag force (Fig. 1C-D), where the inferred friction coefficient showed very little variance within and between fish (Fig. S1B-D). The durations of successive epochs of acceleration ($\sim 200 \pm 104\ ms$) and deceleration ($\sim 250 \pm 160 ms$) were very weakly correlated, and the rate of switching between them was strongly related to the speed of the fish (Fig. S1E-G). We further found that fish made turns mostly during acceleration epochs (Fig. 1E, S1H,I and movie M2). We note that the continuous motion of the adult fish makes these kinematic states very different than the distinct stop-and-go bouts of zebrafish larvae *(45)*.

The segmentation of fish kinematics into clear epochs that have simple functional forms suggests that fish may not be using a universal ongoing computation to determine their behavior at every time instant, as has been suggested previously *(8, 24, 25)*. Furthermore, we find clear anisotropies in group structure, implying that simple distance-based or topology-based models of social interactions, common in the literature, may fall short in explaining individual zebrafish trajectories *(22, 24, 25)*. Specifically, we find that the adult zebrafish prefer to be on the side of other fish (Fig. 1F) within ~1.5 body lengths, and that they typically demonstrated aligned swimming directions when they are directly in front, behind, or on the side of another fish (Fig. 1G).

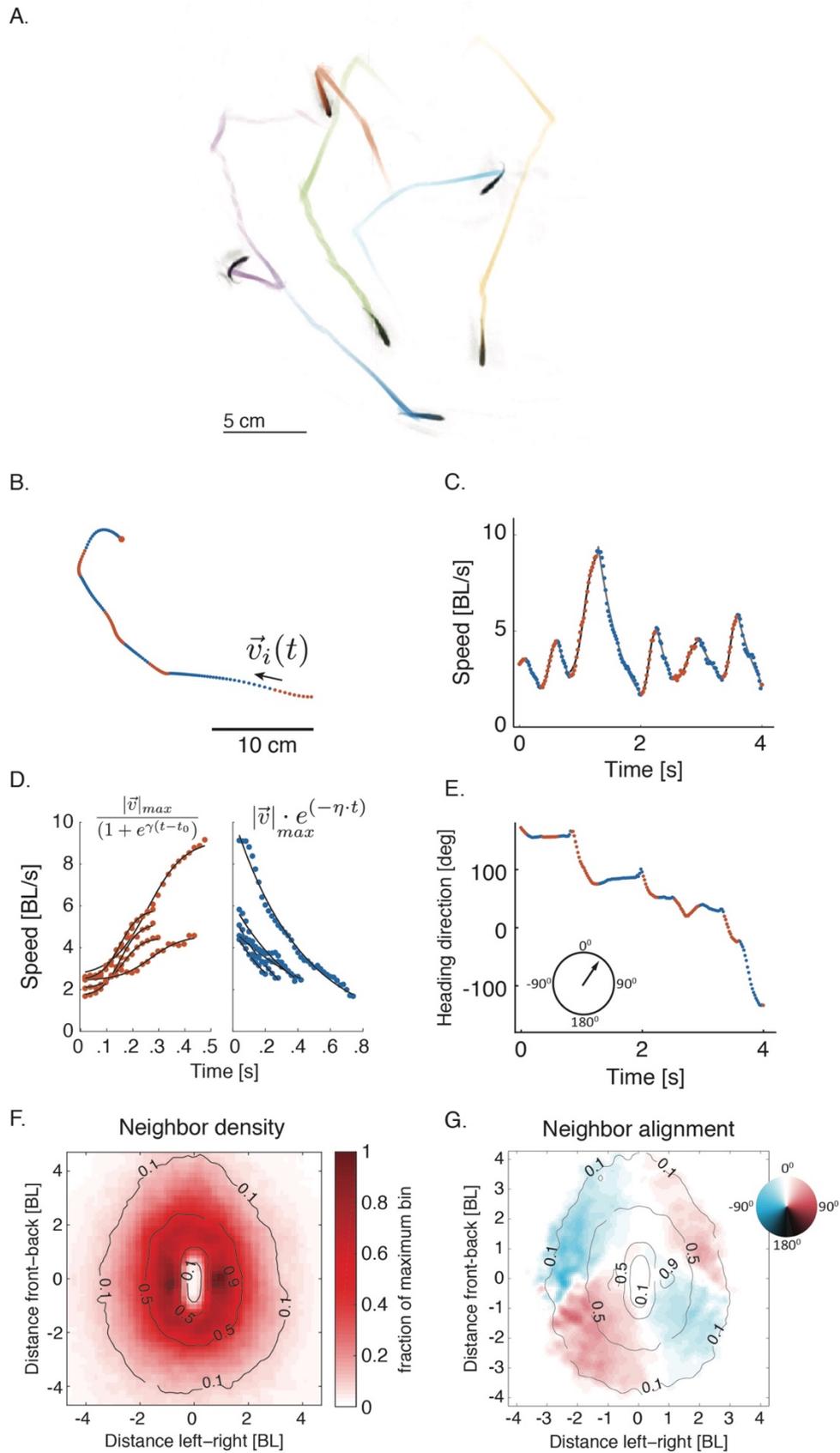

**Fig 1: Kinematic states of individual fish and group structure**. **A.** Snapshot of the tracks of 6 freely swimming fish in a pseudo 2D circular arena. **B.** A short segment of the swimming pattern of a single fish from the group, down-sampled to 50Hz for visualization (dots). Dot color indicate if a fish is accelerating (red) or decelerating (blue) **C.** Speed profile of the trajectory in B. **D.** Functional fits to the acceleration and deceleration epochs in C (see SI Methods). **E.** Heading direction vs. time for the segment shown in B. Directional changes occur predominantly during acceleration epochs (see Fig S1H). **F.** Density map of neighboring fish relative to a focal fish situated at [0,0] pointing north. **G.** Density map of directional alignment of neighboring fish relative to the direction of motion of the focal fish – each point shows the mean alignment value of fish in that bin, with 0 representing perfect alignment (see SI Methods).

We therefore modeled the behavior of individual fish in a group using two modes of information processing: a 'passive' mode where inertia and friction control the movement of the fish, with no sensory or social influence, and an 'active' mode where an additional sensory term, described by a spatio-temporal receptive field (RF) model of sensory and social processing, contributes to the change in velocity. In detail, we discretize time into bins of size $\Delta t$ and denote the measured instantaneous change in velocity of fish $i$ in the group as $\Delta \vec{v}_i(t)$. We model the change in velocity in the passive mode as 'gliding' where water friction slows down the fish (Fig. 2A):

$$\Delta \vec{v}_i^{passive}(t) = -\eta \vec{v}_i(t - \tau_{iner}) \qquad (1)$$

where $\eta$ is the friction coefficient, estimated from fitting deceleration epochs (Fig. 1D and SI Methods), and $\tau_{iner}$ is a short time-constant (chosen here to be 50 ms).

In the active mode, we assume that sensory information and social interactions are taken into account by the fish, and the change in velocity of fish $i$ at time $t$ is given by

$$\Delta \vec{v}_i^{active}(t) = \Delta \vec{v}_i^{passive}(t) + \Delta \vec{v}_i^{RF}(t) \qquad (2)$$

The interaction term $\Delta \vec{v}_i^{RF}(t)$ is given by a spatio-temporal receptive field (RF) model (Fig. 2B):

$$\Delta \vec{v}_i^{RF}(t) = \sum_{j,k} \beta_j(k) \cdot \vec{v}_j(t - k\Delta t) + \sum_{l,k} \beta_l(k) \cdot \vec{d}_l(t - k\Delta t). \qquad (3)$$

The first term is a social interaction term, summing over the past swimming velocities of neighboring fish, where the weights of spatial bin $j$ at time $t - k\Delta t$ are given by $\beta_j(k)$, and $\vec{v}_j(t - k\Delta t)$ is the velocity of the fish in that bin. The second term is the contribution of non-social sensory information, where $\vec{d}_i(t - k\Delta t)$ is a vector tangent to the wall closest to the fish, and $\beta_l(k)$ are the weights associated with that bin. Models were fit on labeled training data, taken from acceleration epochs (see SI Methods); the number of spatial bins and the extent of the temporal history (that together determine the number of parameters) were chosen to maximize model performance using penalized regularization.

The 'passive' and 'active' models give very different predictions for $\vec{v}_i(t)$ at different times along the trajectory of a fish swimming in a group. Fig. 2C shows examples of the different predictions of the two models, on top of a segment of a complex swimming pattern of one fish in a group of three (neighboring fish not shown). Along most of the trajectory, the two models alternate in terms of their accuracy in predicting behavior. Fig. 2D shows the models' prediction errors as a function of time on a short segment of held-out test data, suggesting that the passive model makes smaller errors mostly during decelerations, whereas the active model makes smaller errors mostly during accelerations. This observation was further supported by analyzing complete fish trajectories and multiple groups of the same size recorded independently (N=6-7 for the different sizes): the active model significantly outperformed the passive model in acceleration epochs, while the passive model outperformed the active model in deceleration epochs (Fig. 2E, P<0.0005 for all group sizes, t-test for dependent samples). Learning a separate RF-model for the deceleration epochs did not result in a significant improvement over the passive model (Fig. S2A,B), reasserting that fish show very weak social responses during decelerations. These results indicate that individual fish alternate between two distinct modes of social information processing, which roughly correspond to acceleration and decelerations epochs; in other words, we hypothesize that the kinematic states of the fish are a good indicator for the mode of social information processing.

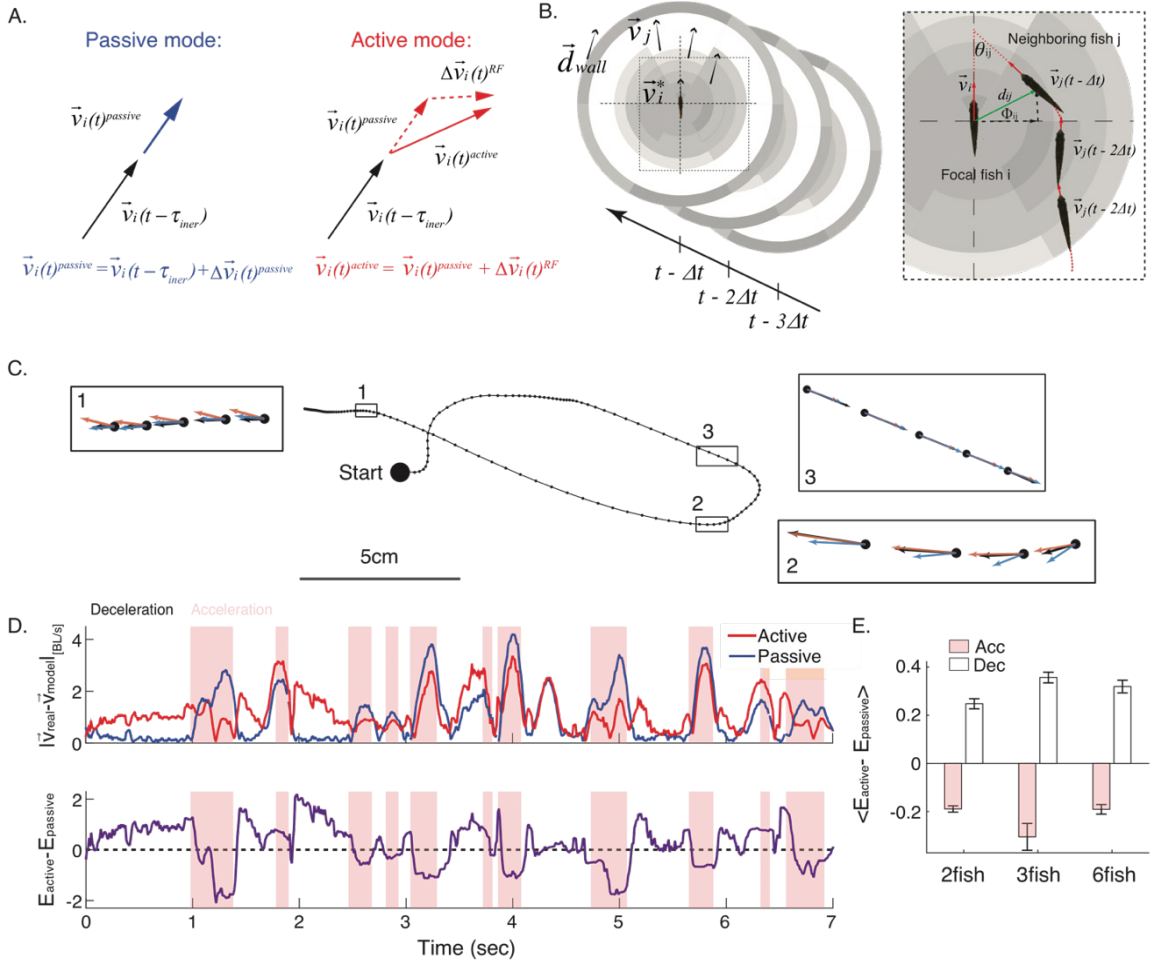

**Fig 2: Modeling fish behavior using active and passive models of social information processing. A**. In the passive model (left), the change in velocity, $\Delta\vec{v}_i(t)$, is given by inertia and friction. In the active model (right), $\Delta\vec{v}_i(t)$ is the sum of the passive component and the contribution of a sensory and social component. **B.** The sensory and social component of the active model is given by a computation based on a spatio-temporal 'receptive field' (RF), where the behavior of conspecifics in spatial 'bins' is weighed with time-dependent parameters (see inset at right and text). **C.** Example of a trajectory of one fish in a group of three, with a comparison of model predictions (passive model in blue, active model in red) and the measured velocity (black). Insets show zoom-in on three representative segments of the trajectory, reflecting the different prediction accuracy of the models at different times. **D.** Top: prediction errors, $E_{model} = |\vec{v}_{real} - \vec{v}_{model}|$, as a function of time for one fish in group of three, using the active mode (red) and the passive mode (blue). Background color designates whether the fish was accelerating (pink) or decelerating (white). Bottom: the difference between the errors of the active and passive models; typically, each of the two models is much more accurate in one of the kinematic states. **E.** Average values of the difference between errors shown in D for groups of 2 fish (N=6 groups), 3 fish (N=7), and 6 fish (N=7); error bars represent SEM.

Since we do not have access to the actual information processing state of the fish, we asked how well can we explain fish behavior if we were to pick the best model for each time point (the one that gives the lowest error when compared to the real velocity). This combined model (see Movie M3) gives an excellent fit to the data both in terms of the speed (Fig. 3A top), and the heading direction of swimming (Fig. 3A bottom). Over all groups, the correlation between the real and the estimated trajectory of the fish was ~0.97 for direction and ~0.94 for speed on test data (Fig. 3B). To further illustrate the importance of the two interleaved modes for describing individual behavior, we compared the accumulated effect of the errors in predicting the instantaneous velocity vectors that each of the models make. Figure 3C shows the 'reconstructed' swimming trajectory of a fish in a group that would result from summing over the instantaneous velocity predictions of each model to obtain a complete trajectory segment (see SI Methods and also Fig. S3C). Repeating this analysis for 5000 3s long segments of a group of 3 fish, we found that combining between the active model of information processing and the passive model (again by choosing the best model at each time point) gave much more accurate reconstructions than either model alone (Fig. 3D left). The reconstruction errors over many trajectory segments for all groups of 3 fish were lower by $37 \pm 5\%$ compared to the passive model alone, and $19 \pm 11\%$ compared to the active model alone; these improvements were similar for groups of 2 and 6 fish (Fig. 3D right, P<0.005 for all group sizes and for both comparisons, t-test for matched samples). Even though the combined model used here is an upper bound for the performance of any mix of the 'active' and 'passive' models, the majority of its performance gains are retained in a model where the kinematic state of the fish is used directly as an indicator of its information processing state (Fig. S3A-E), as hypothesized above.

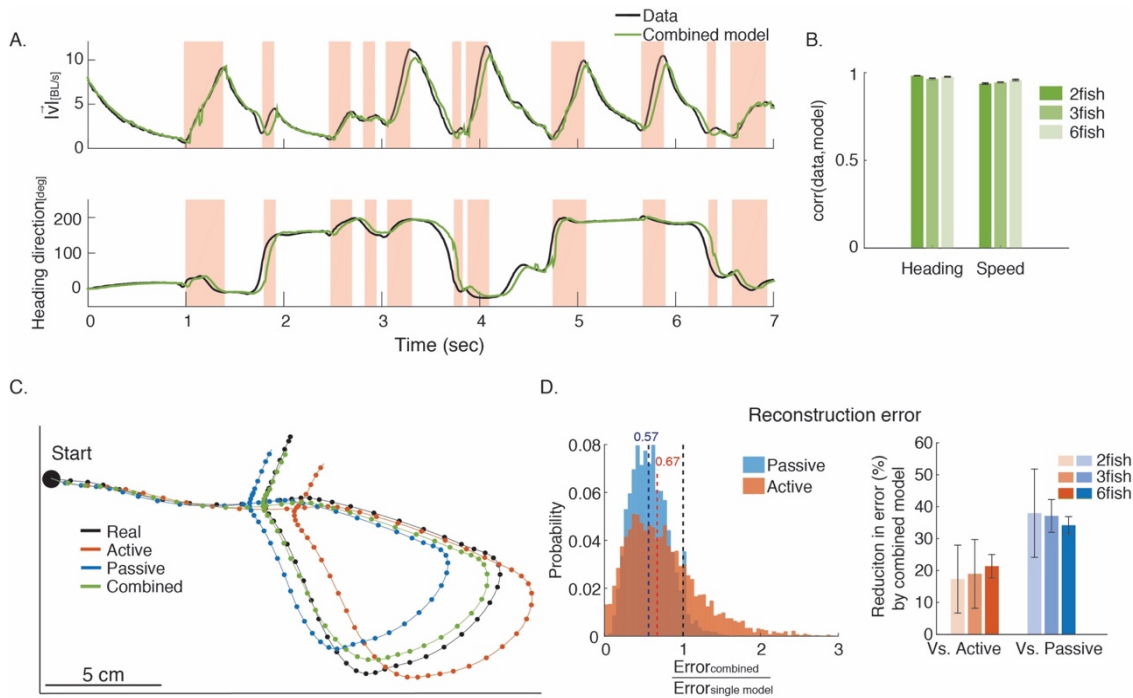

**Fig 3: Accurate prediction of velocity and fish trajectory reconstruction. A.** Examples of the measured speed (top) and heading direction (bottom) of a single fish in a group of 3 (black) and the prediction obtained using the combination of the active and passive models, picking the better model at each time point (green). **B.** Average Pearson's correlation between measured speed and heading direction and model predictions for all fish in groups of the same size (error bars = SEM, N=6,7,7 groups) **C.** A short segment (3s) of the trajectory of a fish in a group of 3 (black) and the reconstructed trajectory obtained by summing the velocity predictions of the active model (red), the passive model (blue), and the optimal combination of the two (green). **D.** Left: Distribution of the ratio between the reconstruction error of the combined model and the reconstruction error of either the active model (red) or the passive model (blue) alone, for 5000 segments similar to the one shown in C. Values below 1 (black dashed line) represent advantage to the combined model. Colored dashed lines represent distribution medians. Right: Median values for the reduction in prediction error of the combined model vs. the active model or the passive model alone (as in the left panel), for groups of different sizes (error bars = STD).

A significant part of the high correlation between model predictions and the data (shown in Fig. 3) originates from the auto-correlation of individual swimming behavior. This is especially true in deceleration epochs, where the correlation between the measured $\vec{v}_i(t)$ and prediction based on $\Delta\vec{v}_i^{passive}(t)$, was $0.986 \pm 0.002$ (see also low prediction error values in Fig. S2A). We therefore focused on the change in velocity that is not explained by autocorrelation and friction. Figure 4A shows the change in velocity that is not explained by the passive component, which we denote as $\Delta\vec{u}_i(t) = \Delta\vec{v}_i(t) - \Delta\vec{v}_i^{passive}(t)$. Clearly, in the deceleration epochs removing the passive component leaves very little change to explain. In the acceleration epochs, the correlation between $\Delta\vec{u}_i(t)$ and the prediction of $\Delta\vec{v}_i^{RF}(t)$ was ~0.5. When we examined the social or sensory contributions to the RF model in isolation, the resulting model's prediction performance was significantly lower than when both information types were included (Fig. 4B, P<0.0005 for all group sizes, t-test for matched samples), with small differences between group sizes (Fig. S4A). The non-additivity of social information and non-social sensory information reflects the redundancy between them. In the current setup, it is impossible to discern whether fish 'read' sensory information about the environment from their own senses, or from the behavior of other fish. We note that the relation between the $\Delta\vec{v}_i$ and the predictions of the models did not indicate a need for a non-linear extension of the (active) RF model (Fig. S4C, cf. LN models in neuroscience; *(46)*). Predicting the entire acceleration epoch using a similar RF model, from the sensory and social information at the beginning of the epoch, performed significantly worse (see SI Methods and Fig. S7).

Our RF model significantly outperformed common models of collective movement in predicting $\Delta\vec{v}_i$, even when the parameters of these competing models were optimized to our data (see SI Methods): we predicted $\Delta\vec{u}_i(t)$ by $\sim 8 \pm 2.5\%$ better on average than a 'zonal model' (24) and $\sim 11.3 \pm 4\%$ better than a 'topological model' (16) (Fig. 4C, P<0.05 for all group sizes and both model comparisons, t-test for matched samples). The advantage of the RF model is even more pronounced when comparing the accuracy of prediction using only social information, as sensory information which is similar across all models obscures part of these differences (Fig. S4B), indicating that the assumptions of the RF-model better match the behavior of swimming zebrafish.

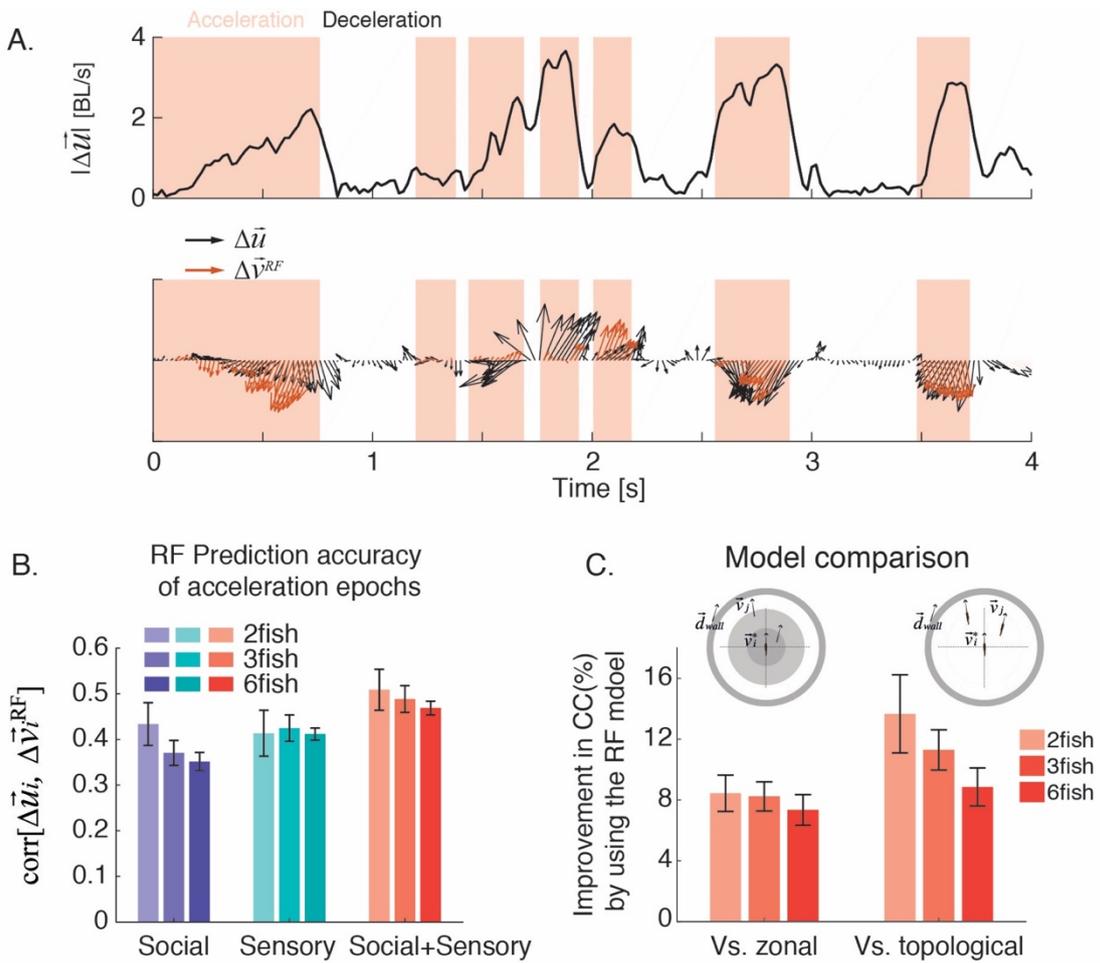

**Fig 4: Active movement changes are accurately predicted by the RF model using both social and sensory information. A.** Top: An example of $|\Delta\vec{u}_i(t)|$, the magnitude of the measured change in velocity after subtracting the passive component (see text) of a single fish in a group of 3 over 4s; background colors mark accelerations (pink) and decelerations (white). Bottom: the corresponding comparison between the measured $\Delta\vec{u}_i(t)$ and the prediction obtained using the RF model (red arrows) in the acceleration epochs. **B.** Prediction accuracy of $\Delta\vec{u}_i(t)$ by RF models that use only the social information component in Eq. 3, only the sensory information component in Eq.

3, or both, for different group sizes (error bars = SEM, N=6,7,7 groups). **C.** Improvement in predicting $\Delta \vec{u}_i(t)$ in acceleration epochs by our RF model relative to its zonal or topological versions (shown in insets; see Methods for details). Improvement values are averaged over all groups of different sizes (error bars represent SEM).

To characterize the spatio-temporal effects of social and sensory information on the movement decisions of a focal fish, we compared the weight maps of the RF models under two different behavioral contexts – fish swimming freely in the arena as described above, and fish who were trained to seek food that is randomly scattered in the arena (see SI Methods). Inhomogeneity in the receptive field map reflect the effects of the relative distance and relative angle of neighbors on the focal fish (Fig. 5A): social effects are strongest in front of the fish and weaker behind it. The weights of the non-social information show the opposite structure, with walls directly to the side of the fish having the strongest effect on its behavior. In general, responses to neighbors are weaker for longer temporal delays, but keep their positive sign. In contrast, the effect of the wall decreases faster with time (see middle weight map in Fig. 5A) and ultimately switches sign. Interestingly, the way fish integrate information from their surroundings changes with the behavioral context (Fig. 5B): effects of arena walls are weaker in food-searching fish, and the effects of fish positioned directly behind the focal fish are positive and stronger (Fig. S5A-B for statistically significant differences between weight maps).

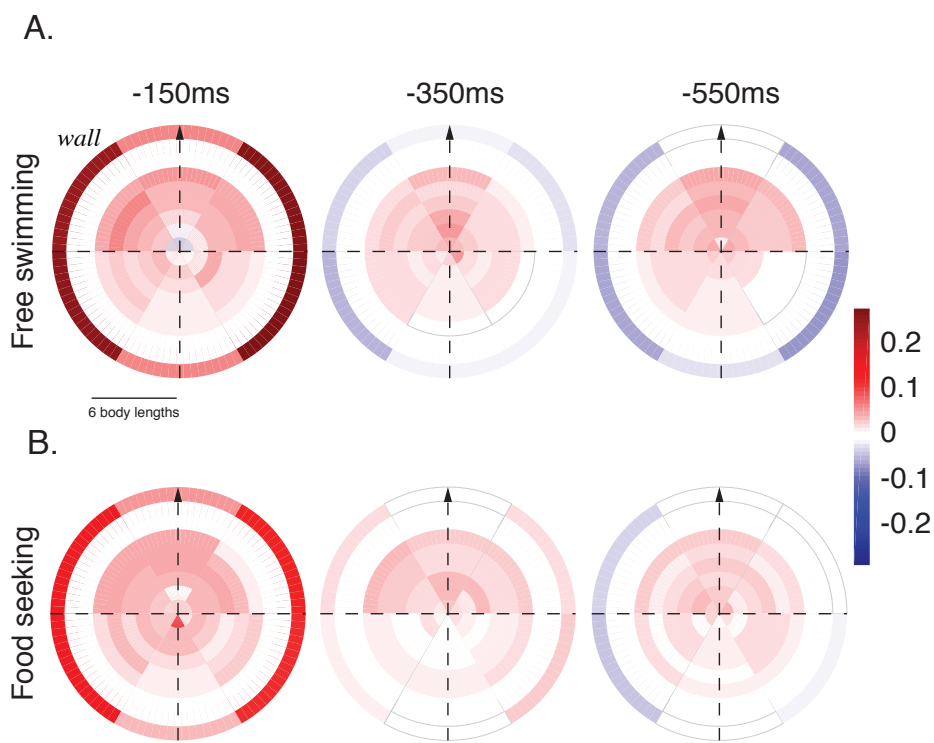

**Fig 5: Receptive field maps show distinct spatial, temporal, and behavioral state dependencies. A.** Receptive fields for a fish in group of 3 fish (N=7), freely swimming in the arena, for the model illustrated in Fig 2B (shown are average $\beta$ values from Eq. 3). Outer circles represent weights associated with the walls of the arena. **B.** Receptive fields obtained for a fish in group of 3 fish (N=7), trained to seek food in the arena (no food is present during the analyzed session).

What does switching between the two modes of information processing at the individual level imply for the behavior of the group? Figures 6A-B show an example of the swimming velocities of 3 fish, decomposed into the speed $s_i(t)$, (Fig. 6A) and the direction of swimming $\theta(\vec{d}_i)$ (Fig. 6B). We asked what are the temporal relations between kinematic states in pairs of fish in the group, by seeking the time lag that would maximize the correlation for short movement segments (1s long) for each fish pair (keeping the identity of the fish throughout the analysis, *(47)* see Methods). The distribution of the time of maximal correlation ($\tau_{max}$) did not show any structure and was indistinguishable from the expectation of fish changing states independently (Fig. 6C); the correlation values also did not differ from what was expected by chance (Fig. S6A). This suggests that the transitions between the two behavioral modes of individual fish are independent. Such organization could give the group a way to sample the sensory space in a distributed and interleaved manner, with no temporal processing gaps, without the need for scheduling. In contrast, analogous analyses identified significant correlations between swimming directions in pairs of fish (Fig. S6A) and a corresponding significant peak in the distribution of temporal lags, suggesting causal relationships (Fig 6D). The absence of statistical dependencies between kinematic states of fish in a group and the presence of dependencies for swimming direction was corroborated by estimating the probability of synchronized states among the fish in the group: i.e. the probability to find $k$ out of the $N$ fish in the group to be accelerating synchronously (Fig. 6E), and the probability of $k$ fish to swim in a similar direction (Fig. 6F). For synchronous accelerations, the probability distribution was symmetric and matched closely the expected distribution if fish were switching states independently of one another (Fig. 6E). The distribution of number of fish swimming synchronously in the same direction had a clear structure and was very different from the expectation for independent fish (Fig. 6F). Thus, independent switching between modes of information processing in individual fish on a time scale of several seconds is consistent with the emergence of correlated directional behavior with clear temporal ordering at the group level.

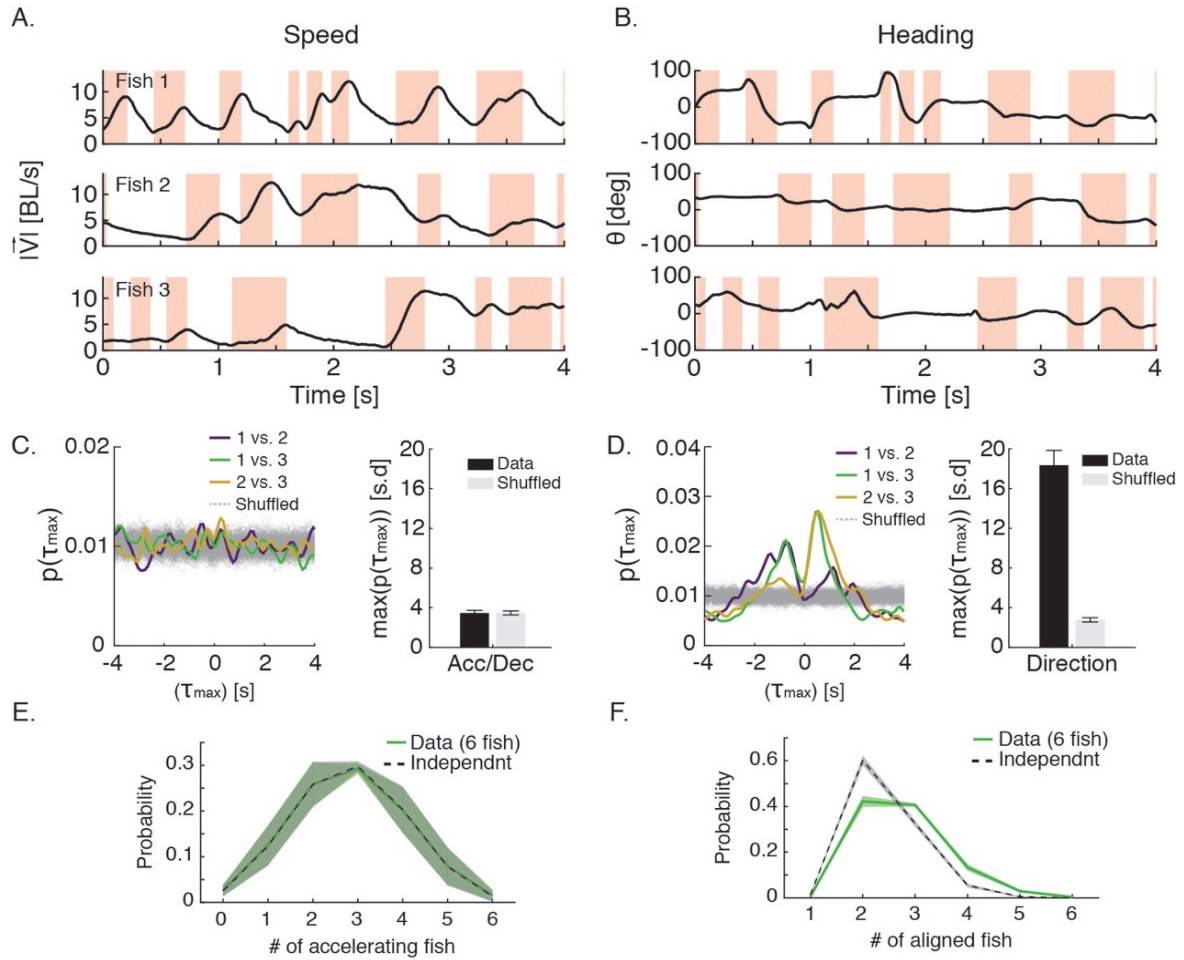

**Fig 6: Asynchronous switching between information processing modes among individual fish in a group and synchronous heading directions of group members. A.** Example of the simultaneous acceleration (pink)/deceleration (white) profiles of 3 fish in a group. **B**. Same as in A, but for the heading direction of the fish. **C.** Left: Distributions of delay time ($\tau_{max}$) that gave the maximum correlations between pairs of fish over 1s long windows, in a group of 3 (colored lines) and shuffled controls (light gray - see SI Methods). Right: Peak correlation value in units of standard deviation of the shuffled data averaged for all pairs of fish within all groups of 3 fish (N=6, P=0.75, t-test for matched samples). **D.** Same as in C, but for the cross-correlation of direction of motion of pairs of fish. Here directional correlation show clear structure and peak times (N=6, P<0.0005, t-test for matched samples). **E.** Average empirical probability distribution $P(a_1, a_2, ..., a_n)$ of the kinematic states of the fish (acceleration or deceleration), where $a_i$ represent the state of fish $i$ (solid green), and the prediction of a model assuming independence between fish $p(a_1)p(a_2)...p(a_n)$ (dashed grey line); light shadings represent SEM. **F.** Same as in E, only for the distribution of heading direction $P(d_1, d_2, ..., d_n)$, where $d_i$ is the direction of fish $i$ discretized into 6 even sized bins (see Methods) (solid green), and the distribution obtained under the assumption of independence $p(d_1)p(d_2)...p(d_n)$.

**Discussion**

Predicting individual behavior of fish in a group, by combining active and passive models of sensory and social information processing, proved to be highly accurate, outperforming commonly used models that assume a universal ongoing computation by individuals. Specifically, spatio-temporal receptive fields captured the computation that a fish performs, surpassing current models that assume simple topological or metric based computation. Moreover, a comparison between food seeking vs. free swimming behavior revealed that the computation employed by the fish depends strongly on context. At the group level, the behavioral modes of individuals seem temporally independent among fish, yet signatures of collective behavior still arise. The approach we presented here merges two distinct lines of inquiry of animal behavior: studies of single-animal behavior that have shown 'discrete behavioral modes' *(36–38)*, and group behavior models that have focused on qualitatively capturing complex collective behavior *(8, 24, 25, 33)* emerging in groups of simple interacting individuals described by a single behavioral mode. Our results show that individual behavioral modes: (i) have clear kinematic proxies, (ii) suggest distinct information processing/computation modes in individuals, and (iii) have a significant impact on group behavior. Beyond an improved model for individual behavior in a group, our approach portrays the group as a collection of diverse individuals whose computations seem temporally discrete and context-dependent, with interactions that are dynamic in space and time.

The model presented here can be improved in several ways. One possibility is to further optimize spatiotemporal filters used to describe the visual field of a fish and to add non-linear components to the prediction model. Improved accuracy would allow us to explore the limits of the computation of individuals and study the implications of noise (sensory or motor) on behavior. Another interesting possibility would be to capture additional aspects of individual and social computations: First, a finer dissection of individual behavior into multiple behavioral modes might reveal further intricate processing. Second, one could define and quantify the transitions between behavioral states in individuals and their dependence on internal factors, as well as social ones. Third, the differences between the receptive fields inferred under different behavioral contexts reflect a dynamic and possibly learned nature of these receptive fields. Modeling how individual fish use different computations based on 'personal' tendencies, past experience, or current needs would bring us closer to dissecting idiosyncratic behavior and understanding its effect at the group level.

The approach we presented here can be readily extended to other animal groups. Moreover, it could be used for exploring different aspects of fine motor behavior and group traits. For example, for fish this could entail mapping their exact visual stimuli *(13)* to tailbeats, which would enable the study of the mapping of sensory and social information into action, possibly in closed-loop experimental settings. Finally, combining our approach with recording of neural activity in members of the group *(48)*, would allow for direct study of social and sensory integration and processing at behavioral and neuronal levels simultaneously.

**Acknowledgements**

We thank Ehud Karpas, Oren Forkosh, Iain Couzin, Ofer Feinerman, and Yadin Dudai for helpful discussions and support. This work was supported by Human Frontier Science Program, European Research Council grant # 311238, an Israel Science Foundation grant #1629/12, as well as research support from Martin Kushner Schnur; and Mr. and Mrs. Lawrence Feis; ES is the Joseph and Bessie Feinberg professorial chair

**References:**

1. Shklarsh A, Ariel G, Schneidman E, Ben-Jacob E (2011) Smart Swarms of Bacteria-Inspired Agents with Performance Adaptable Interactions. PLoS Comput Biol 7(9):e1002177.
2. Reid CR, et al. (2016) Decision-making without a brain: how an amoeboid organism solves the two-armed bandit. J R Soc Interface 13(119):20160030.
3. Perna A, et al. (2012) Individual Rules for Trail Pattern Formation in Argentine Ants (Linepithema humile). PLoS Comput Biol 8(7):e1002592.
4. Gelblum A, et al. (2015) Ant groups optimally amplify the effect of transiently informed individuals. Nat Commun 6:7729.
5. Greenwald E, Segre E, Feinerman O (2015) Ant trophallactic networks: simultaneous measurement of interaction patterns and food dissemination. Sci Rep 5:12496.
6. Herbert-Read JE, et al. (2011) Inferring the rules of interaction of shoaling fish. Proc Natl Acad Sci 108(46):18726–18731.
7. Katz Y, Tunstrøm K, Ioannou CC, Huepe C, Couzin ID (2011) Inferring the structure and dynamics of interactions in schooling fish. Proc Natl Acad Sci 108(46):18720–18725.
8. Gautrais J, et al. (2012) Deciphering Interactions in Moving Animal Groups. PLoS Comput Biol 8(9):e1002678.
9. Miller N, Gerlai R (2012) From Schooling to Shoaling: Patterns of Collective Motion in Zebrafish (Danio rerio). PLoS ONE 7(11):e48865.
10. Tunstrøm K, et al. (2013) Collective States, Multistability and Transitional Behavior in Schooling Fish. PLoS Comput Biol 9(2):e1002915.
11. Berdahl A, Torney CJ, Ioannou CC, Faria JJ, Couzin ID (2013) Emergent Sensing of Complex Environments by Mobile Animal Groups. Science 339(6119):574–576.
12. Arganda S, Pérez-Escudero A, Polavieja GG de (2012) A common rule for decision making in animal collectives across species. Proc Natl Acad Sci 109(50):20508–20513.
13. Rosenthal SB, Twomey CR, Hartnett AT, Wu HS, Couzin ID (2015) Revealing the hidden networks of interaction in mobile animal groups allows prediction of complex behavioral contagion. Proc Natl Acad Sci 112(15):4690–4695.
14. Nagy M, Ákos Z, Biro D, Vicsek T (2010) Hierarchical group dynamics in pigeon flocks. Nature 464(7290):890–893.
15. Bialek W, et al. (2012) Statistical mechanics for natural flocks of birds. Proc Natl Acad Sci 109(13):4786–4791.
16. Ballerini M, et al. (2008) Interaction ruling animal collective behavior depends on topological rather than metric distance: Evidence from a field study. Proc Natl Acad Sci 105(4):1232–1237.
17. Shemesh Y, et al. (2013) High-order social interactions in groups of mice. eLife 2:e00759.

18. Strandburg-Peshkin A, Farine DR, Couzin ID, Crofoot MC (2015) Shared decision-making drives collective movement in wild baboons. Science 348(6241):1358–1361.
19. Faria JJ, Dyer JRG, Tosh CR, Krause J (2010) Leadership and social information use in human crowds. Anim Behav 79(4):895–901.
20. Moussaïd M, Helbing D, Theraulaz G (2011) How simple rules determine pedestrian behavior and crowd disasters. Proc Natl Acad Sci U S A 108(17):6884–6888.
21. Vicsek T, Czirók A, Ben-Jacob E, Cohen I, Shochet O (1995) Novel Type of Phase Transition in a System of Self-Driven Particles. Phys Rev Lett 75(6):1226.
22. D'Orsogna MR, Chuang YL, Bertozzi AL, Chayes LS (2006) Self-Propelled Particles with Soft-Core Interactions: Patterns, Stability, and Collapse. Phys Rev Lett 96(10):104302.
23. Couzin ID, Krause J, Franks NR, Levin SA (2005) Effective leadership and decision-making in animal groups on the move. Nature 433(7025):513–516.
24. Huth A, Wissel C (1992) The simulation of the movement of fish schools. J Theor Biol 156(3):365–385.
25. Couzin ID, Krause J, James R, Ruxton GD, Franks NR (2002) Collective memory and spatial sorting in animal groups. J Theor Biol 218(1):1–11.
26. Bonabeau E, Dorigo M, Theraulaz G (2000) Inspiration for optimization from social insect behaviour. Nature 406(6791):39–42.
27. Branson K, Robie AA, Bender J, Perona P, Dickinson MH (2009) High-throughput ethomics in large groups of Drosophila. Nat Meth 6(6):451–457.
28. Nathan R, et al. (2012) Using tri-axial acceleration data to identify behavioral modes of free-ranging animals: general concepts and tools illustrated for griffon vultures. J Exp Biol 215(Pt 6):986–996.
29. Ákos Z, Beck R, Nagy M, Vicsek T, Kubinyi E (2014) Leadership and Path Characteristics during Walks Are Linked to Dominance Order and Individual Traits in Dogs. PLOS Comput Biol 10(1):e1003446.
30. Nagy M, et al. (2013) Context-dependent hierarchies in pigeons. Proc Natl Acad Sci 110(32):13049–13054.
31. Khuong A, et al. (2016) Stigmergic construction and topochemical information shape ant nest architecture. Proc Natl Acad Sci U S A 113(5):1303–1308.
32. Seeley TD, et al. (2012) Stop Signals Provide Cross Inhibition in Collective Decision-Making by Honeybee Swarms. Science 335(6064):108–111.
33. Huth A, Wissel C (1994) The simulation of fish schools in comparison with experimental data. Ecol Model 75–76(0):135–146.
34. Zienkiewicz A, Barton DAW, Porfiri M, Bernardo M di (2014) Data-driven stochastic modelling of zebrafish locomotion. J Math Biol:1–25.
35. Parrish JK, Edelstein-Keshet L (1999) Complexity, Pattern, and Evolutionary Trade-Offs in Animal Aggregation. Science 284(5411):99–101.
36. Girdhar K, Gruebele M, Chemla YR (2015) The Behavioral Space of Zebrafish Locomotion and Its Neural Network Analog. PLoS ONE 10(7):e0128668.
37. Stephens GJ, Johnson-Kerner B, Bialek W, Ryu WS (2008) Dimensionality and Dynamics in the Behavior of C. elegans. PLOS Comput Biol 4(4):e1000028.

38. Berman GJ, Choi DM, Bialek W, Shaevitz JW (2014) Mapping the stereotyped behaviour of freely moving fruit flies. J R Soc Interface 11(99):20140672.
39. Suriyampola PS, et al. (2015) Zebrafish Social Behavior in the Wild. Zebrafish 13(1):1–8.
40. Portugues R, Feierstein CE, Engert F, Orger MB (2014) Whole-Brain Activity Maps Reveal Stereotyped, Distributed Networks for Visuomotor Behavior. Neuron 81(6):1328–1343.
41. Aoki T, et al. (2013) Imaging of Neural Ensemble for the Retrieval of a Learned Behavioral Program. Neuron 78(5):881–894.
42. Bianco IH, Engert F (2015) Visuomotor Transformations Underlying Hunting Behavior in Zebrafish. Curr Biol 25(7):831–846.
43. Ahrens MB, et al. (2012) Brain-wide neuronal dynamics during motor adaptation in zebrafish. Nature 485(7399):471–477.
44. Buske C, Gerlai R (2011) Shoaling develops with age in Zebrafish (Danio rerio). Prog Neuropsychopharmacol Biol Psychiatry 35(6):1409–1415.
45. Müller UK, Leeuwen JL van (2004) Swimming of larval zebrafish: ontogeny of body waves and implications for locomotory development. J Exp Biol 207(5):853–868.
46. Chichilnisky EJ (2001) A simple white noise analysis of neuronal light responses. Netw Bristol Engl 12(2):199–213.
47. Pérez-Escudero A, Vicente-Page J, Hinz RC, Arganda S, de Polavieja GG (2014) idTracker: tracking individuals in a group by automatic identification of unmarked animals. Nat Methods 11(7):743–748.
48. Vinepinsky E, Donchin O, Segev R (2017) Wireless electrophysiology of the brain of freely swimming goldfish. J Neurosci Methods 278:76–86.

## Supplementary information

## Methods:

### Experimental system.
75 adult zebrafish (Danio Rerio), purchased from Aquazone Israel LTD, at approximately 1:1 male: female ratio were studied. Fish were housed separately in their designated groups for at least one month prior to experiments. Environmental conditions were constant using a re-circulating system and multistage filtration, with water temperature of 27-28°C, conductivity of 600-700 $\mu siemens$ and PH levels of 7.7–8. Lighting was kept at 14:10 light/dark cycle with fluorescent lights. Unless otherwise indicated, fish were fed twice a day a mixture of dry flake food. Experimental arena consisted of a large 1.2m over 1.1m rectangular aquarium with circular arenas of different diameters placed in it (see Movie M1). Water levels were kept at a depth of about 5 cm to constitute a pseudo 2-dimensional environment.

Video recording was done using an industrial recording system with a Vieworks VC-2MC-M340 digital camera with an 8mm lens, connected to a Karbon-CL frame grabber. Camera was attached to the ceiling over the test aquarium approximately 150cm above water level to capture the entire arena.

### Data extraction.
Videos were analyzed off-line to extract the physical properties of the fish (size, position, orientation). Position data was used to estimate fish trajectories using a designated tracker. All image processing and tracking was done using Matlab with software written in our lab. Briefly, fish were first detected as darker blobs over the lighter background, and their physical properties calculated. Next, the center of mass of fish were connected frame by frame to give the estimated track of each fish. When several fish were close to one another an additional step was taken to estimate the most likely number of fish in the large blob and their centers. This tracking method provided accurate detection of fish in >90% of the frames analyzed, but did not ensure constant identities of the fish. When needed, fish identities were corrected using the IdTracker software (1). Fish trajectories were smoothed using a Savitzky-Golay filter (2) spanning 33 frames which constitutes ~1/3 of a second. Fish positions were set as the coordinates of the fish center: $\vec{c}_i(t) = [x(t)_i, y(t)_i]$, and fish velocity was estimated as the difference between two consecutive points: $\vec{v}(t)_i = \vec{c}(t)_i - \vec{c}(t-1)_i$. Direction of motion of the fish was defined as $\vec{d}(t)_i = \frac{\vec{v}(t)_i}{|\vec{v}(t)_i|}$, or $\theta(t)_i$ as the angle of $\vec{d}(t)_i$, and angular velocity was given by $\omega(t)_i = \theta(t)_i - \theta(t-1)_i$. As fish tend to respond strongly to walls of the arena, we did not use data from fish positioned 'close' to the boundary - all distances smaller than the median of the wall distance distribution were discarded from further analysis.

### Behavioral experiments.
Free-swimming. Prior to behavioral experiments, fish were habituated to the circular arena (95cm diameter) for short sessions ~10 minutes for 2 days. We then filmed their free-swimming behavior for 30 or 60 minutes.

Food-seeking. To train fish to seek for food in the arena we conducted a 7-day training protocol. On each day, fish were transferred from their home tank to the experimental arena where a constant number of flakes (~4mm in diameter) were scattered randomly on the water surface. Fish were allowed 5 minutes to consume the flakes and then netted and returned to their home tanks. To facilitate learning, the effective size of the arena used by the fish was increased over days (from a small box of 25X25cm to circular arenas of diameter of 57cm, 82cm, 95cm). During training, no other food was given to the fish, and on days 6-7 fish were deprived of food. On day 8, fish were transferred again to the test tank and their behavior was recorded, with no food present.

**Kinematic models of fish acceleration and deceleration segments.**
Segmentation of speed profiles into acceleration and deceleration epochs was done by detecting the minima and maxima of the speed profile using numerical differentiation. To overcome local noise, we constrained two local minima to be separated by at least 4 frames (or 40 ms) and the single highest extrema between two such minima points was taken as the local maximum point (and end of acceleration). Each segment $j$ of fish $i$ was fitted individually: a single exponential $s_i^{dec[j]}(t) = max\,(s_i^{dec[j]}) \cdot e^{(-\eta_{ij} \cdot t)}$ was used to fit decelerations and a sigmoid $s_i^{acc[j]}(t) = \frac{max\,(s_i^{acc[j]})}{1+e^{\gamma_{ij}(t-t_0)}}$ for accelerations. Maximum speeds for both models - $max\,(s_i^{acc[j]})$ and $max\,(s_i^{dec[j]})$ were obtained form the data as was the midpoint of the sigmoidal function $t_0$, leaving the 'slope' of both models $\eta_{ij}$ and $\gamma_{ij}$ as free parameters to be fitted (see Fig. S1B-D). Fitting was done only for segments that were at least 100ms long (i.e. having at least 10 data points for fitting).

**Neighbor maps.**
Neighbor position maps were based on 2-dimensional histograms of neighbors' positions in space, relative to a focal fish whose orientation was defined as "north". Histograms were smoothed using a rotationally symmetric Gaussian low pass filter ($\sigma^2 = 0.13$ [body lengths]).
To estimate the neighbor alignment maps, the average direction of motion of fish in each bin was calculated using all frames in which the bin was occupied. The angular deviation of that average vector from the direction of motion of the focal was expressed in angles [-180 180] where positive values represent deviation to the right ('east') and negative values represent deviation to the left ('west').

**Model Fitting.**
Receptive field models were fit using a Lasso least-squares regression (3) with cross validation. Briefly, for a given non-negative $\lambda$ we calculated

$$\min_{\vec{\beta}} \left\{ \frac{1}{2N} \sum_{t=1}^{N} (\vec{v}_t - \vec{x}^T \cdot \vec{\beta})^2 + \lambda \sum_{j=1}^{p} |\beta_j| \right\}, \qquad (1)$$

where $\vec{v}_t$ is the empirical velocity of fish in time instance $t$, $\vec{x}$ is a set of velocities of neighboring fish in the spatio-temporal receptive field around the fish (see Eq. (3) in main text), $\vec{\beta}$ is a vector of model parameters or bin weights (of length p), and N is the total number of observations. Repeating this procedure for different values of $\lambda$, we found the set of parameters $\vec{\beta}$ that minimized the cross-validation error for held out data. This regularization process usually reduces the effective number of parameters (setting some of the weights to zero) resulting in a sparser model.

**Competing models.**
We have compared our receptive-field parameterization of space, which depends on both angle and distance of neighbors from the focal fish (Fig. 2B), to two commonly used parameterization of space: a zonal model (4–6) where only the distance of neighboring fish is taken into account, and a topological model (7) where neighbors are weighed according to their topological order (first neighbor, second neighbor, etc) ignoring their metric distance (see insets in Fig. 4C for model sketch). Mathematically we get a similar expression as in eq. (3) in the main text

$$\Delta \vec{v}_i^{Model}(t) = \sum_{j,k} \beta_j(k) \cdot \vec{v}_j(t - k\Delta t) + \sum_{l,k} \beta_l(k) \cdot \vec{d}_l(t - k\Delta t) \qquad (2)$$

where the first term is the social interaction term, but spatial weights are assigned to bins according to their distance from the focal fish (zonal model) or to topological neighbors. Temporal binning was kept similar to that used in the RF model for comparison.

**Model parameters.**
Discretization of the receptive field into spatio-temporal bins and temporal bins of the topological and zonal models were chosen using a non-exhaustive search of parameter space, whereas zonal model radii, were optimized for the cases of either 2 or 3 rings. Thus, the RF parameters used here give only a lower bound on the accuracy of this model since an exhaustive search of the parameter space can optimize the obtained results. All parameters used in the models are listed below:

For the RF-model:

| Parameter name | Meaning | Value used in analysis |
| --- | --- | --- |
| # of sectors | Number of sectors in the Receptive field | 6 sectors |
| Start angle | The middle point of the first sector with respect to fish motion. | 0 degrees (north) |
| # of rings | Number of rings in the receptive field. | 6 rings |
| Ring size | $R_{i+1} - R_i$ where $R_i$ is the radius of ring i. | 1 body length |

For the Zonal models

| Parameter name | Meaning | Value used in analysis |
| --- | --- | --- |
| # of rings | Number of rings in the zonal model | 2 rings |
| Ring size | $R_{i+1} - R_i$ where $R_i$ is the radius of ring i. | 3 body length |

Temporal parameters similar for all models:

| | | |
| --- | --- | --- |
| # of steps back | Number of steps back in time used for prediction | 3 steps |
| $\Delta t_1$ | Time window between fish motion and neighbor first neighbor response configurations | 150 ms (15 frames) |
| $\Delta t_{2,3}$ | Time window between first and second response configurations, and between second and third. | 200 ms (20 frames) |

**Trajectory reconstruction.**

Predicted positions of fish $i$, $C_i(t)$ over time, were obtained using the instantaneous velocities $\vec{v}_i(t)$ predicted by the models where $\vec{C}_i(t) = \vec{C}_i(t-1) + \vec{v}_i(t)$; $\vec{v}_i(t)$ were obtained either by the Active model or the Passive model (see main text). The 'combined model' is where for each time step t we use the prediction of the model that gave the lower error $E_{model} = |\vec{v}_{real} - \vec{v}_{model}|$ between the real and predicted velocities for <u>that</u> time step. We note that this is only a locally 'optimal' choice, that does not guarantee that the total error between real and predicted trajectories will be minimal.

**Directional and behavioral cross-correlation analysis.**

Time windowed directional cross-correlation (8, 9) was estimated using a short window of the response of fish $i$ ($L = 1s$), and the responses of fish $j$ over a range $-4 \leq \tau \leq 4$ sec (in 10ms increments):

$$C_{ij}(t) = \sum_\tau \vec{v}_i(t)\, \vec{v}_j(t-\tau)$$

where $\vec{v}_i$ and $\vec{v}_j$ are fish velocities. The time delay $\tau$ that gives the maximal correlation for this segment is then considered to be the temporal relation between animals. We then calculate $C(t)$ for all time points and for all pairs of fish in a given group and find the maximal correlation for each segment $C_{ij}^{max}$ (Fig. S6A), and the temporal delay corresponding to this maximal correlation $\tau_{max(c)}$. Histograms of temporal relations (Fig 6D and Fig. S6B bottom) are then constructed using these $\tau_{max(c)}$ values, keeping only temporal delays of high correlations $C_{ij}^{max} > 0.9$, to focus only on instances where the fish are actually responding to one another and to ensure that more that 10% of all correlations are retained both for the actual data and for the shuffled analysis (see below). Importantly, choosing different thresholds for $C_{ij}^{max}$ did not qualitatively change the results obtained (Fig. S6B, bottom).

Behavioral state correlation was estimated in a similar manner, but in this case, we use $x_i$ and $x_j$ which are binary variables denoting acceleration (1) or decelerations (0) (Fig S6A Left). When constructing the temporal delay distributions (Fig. 6C) we kept $\tau_{max(c)}$ values corresponding to $C_{ij}^{max} \geq 0.8$ for the behavioral state correlation, as to retain at least 10% of the correlation in the shuffled analysis. Choosing different threshold values for this analysis as well gave similar results (Fig. S6B top).

To obtain a null distribution of expected correlation values we repeated the same time-windowed cross-correlation analysis as formulated above for both direction and behavioral state, but with each fish's trajectories randomly shifted in time. We repeated this full analysis 1000 times for each pair of fish (see gray lines in Fig 6C,D and in Fig. S6A,B). To compare the magnitude of the peaks of the time delay distributions $max(P(\tau_{max}))$ found in real pairs and that of the time-shuffled pairs, we normalized the maximal values of $P(\tau_{max})$ using the mean and standard deviation of the maxima found in the repetitions of the shuffled analysis

$$score = \frac{peak - \mu_{shuffled}}{\sigma_{shuffled}}$$

where $\mu_{shuffled}$ and $\sigma_{shuffled}$ are the mean and standard deviation of the peak in the shuffled data. This why we are effectively conducting a bootstrap analysis comparing the maximal probability in our data to that found in the shuffled analysis (Fig 6C,D right).

**Estimating empirical distributions of collective states.**
We evaluated the joint probability distributions of synchronous fish states, $P(a_1, a_2, ..., a_n)$, where $a_i$ is the kinematic state of fish $i$ (we set 1 for accelerations and 0 for decelerations) and calculate the probability of seeing either $1, 2, ..., k$ out of $n$ fish occupying the same state. For comparison, we estimate the independent probability distribution, $p(a_1)p(a_2)...p(a_n)$, where $p(a_i)$ is the independent probability of fish $i$ to be in each state, taken as the average over the entire session.

Similarly, we estimated the probability distribution of the synchronous swimming direction of fish in a group as $P(d_1, d_2, ..., d_n)$, where $d_i$ is the direction of swimming of fish $i$ binned into 6 angular even sized bins. We compared it to the distribution of directions obtained under a similar independence assumption, $p(d_1)p(d_2)...p(d_n)$.

**Predicting full acceleration epochs.**
We define the response of a fish over a full acceleration epoch as the integral of the speed $\int_0^t S_i dt$ and of the angular velocity $\int_0^t \omega_i dt$ (see Fig S7A). Angular velocity is taken to be $\omega = \frac{d\theta}{dt}$ where $\theta$ is the heading direction of the fish. Fitting an RF model to predict these quantities was a similar process to the one used for velocity prediction. For comparison, we also fitted the RF-model for instantaneous velocities again, reducing the amounts of the data used in the fitting process so it would match the number of complete acceleration epochs. To this end, we chose a single representative velocity positioned 150ms after the transition between deceleration and acceleration. This choice gave very similar results to the RF-model fits on the entire data set (see Fig S7C).

**SI Figures:**

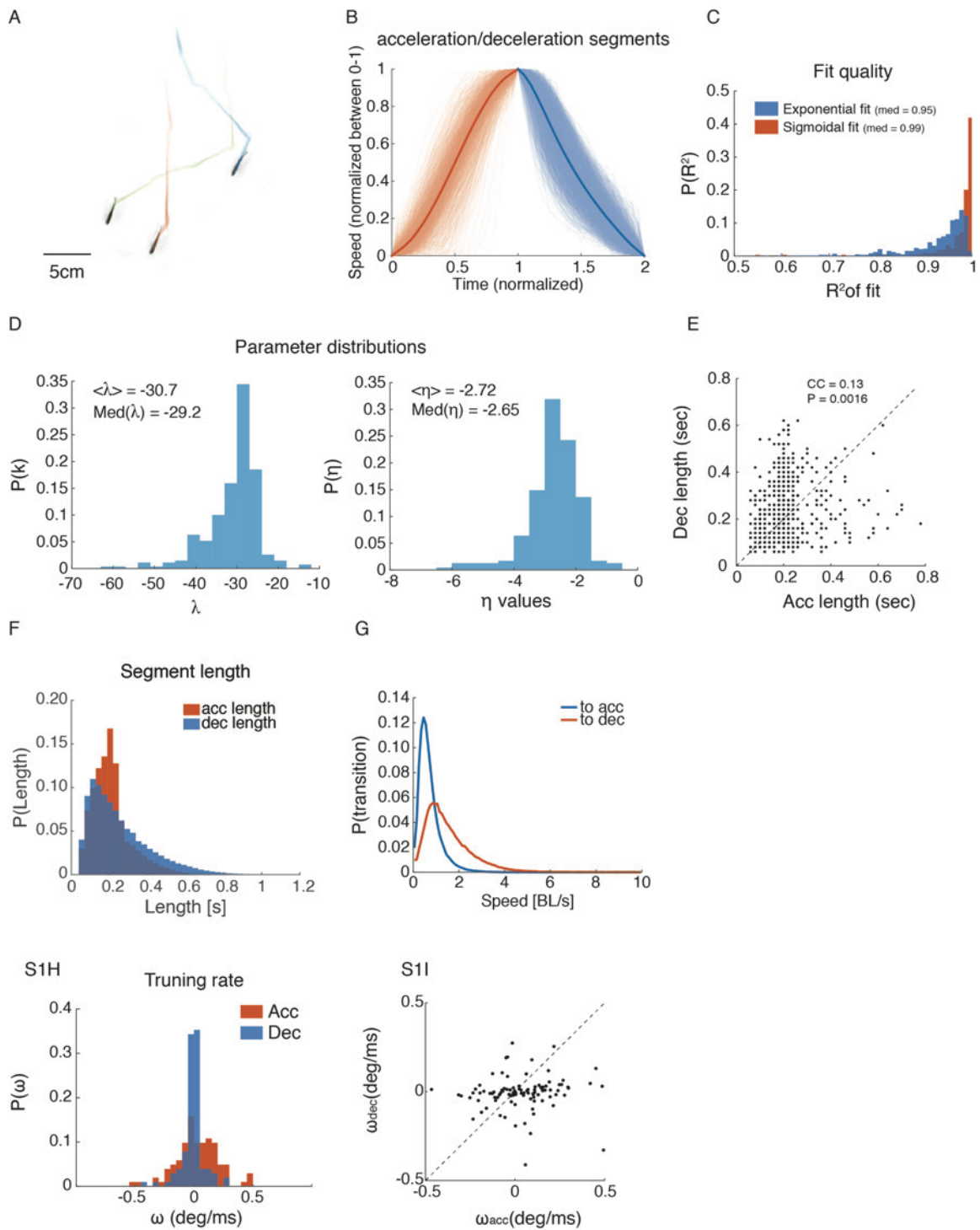

**Figure S1: Kinematic states of individual fish in a group. A.** A snapshot of the tracks of 3 fish. **B.** All acceleration and deceleration epochs for a single fish over 5 minutes of swimming, with time normalized between zeros and 2 (1 marks the transition from acceleration to deceleration) and speed normalized between zeros and one. Bold lines are the means over all epochs. **C.** Distributions of $R^2$ values for all the segments shown in B, with a median value of 0.99 for accelerations and 0.95 for decelerations representing high model fit accuracies. **D.** Fitted parameters for accelerations (left) and decelerations (right). Sigmoid slope $\gamma$ shows a distribution peaked around $-30\ [\frac{1}{10ms}]$, and friction coefficient around $\eta = -2.7[\frac{1}{10ms}]$. **E.** Duration of successive acceleration and deceleration epochs (shown in B) plotted one against the other, shows a very week linear relationship between the two (Pearson's $CC = 0.13, P = 0.0016$). **F.** Distributions of acceleration and deceleration durations for all fish in all groups. Acceleration epochs had a mean duration of $\sim 200 \pm 104\ ms$, and were generally shorter than decelerations with a mean duration of $\sim 250 \pm 160 ms$ **G.** Probability of switching states as a function of swimming speed. As can be expected, the probability of switching from deceleration to acceleration is higher for low speed values with a peak at 0.45 BLs/s and a narrow distribution (middle 95% of the distribution between 0.05-1.85 BLs/s), while switching back to deceleration is peaked at 1 BLs/s with a much wider distribution (middle 95% of the distribution between 0.2-4.15 BL/s) **H.** Distributions of the angular velocity $\omega = \frac{d\theta}{dt}$ with $\theta$ being the heading of the fish, for the acceleration and deceleration epochs shown in B. Note the narrow distribution centered around zero for decelerations. **I.** Angular velocity $\omega$ of successive acceleration and deceleration epochs. No clear relationship seems to exist between successive epochs.

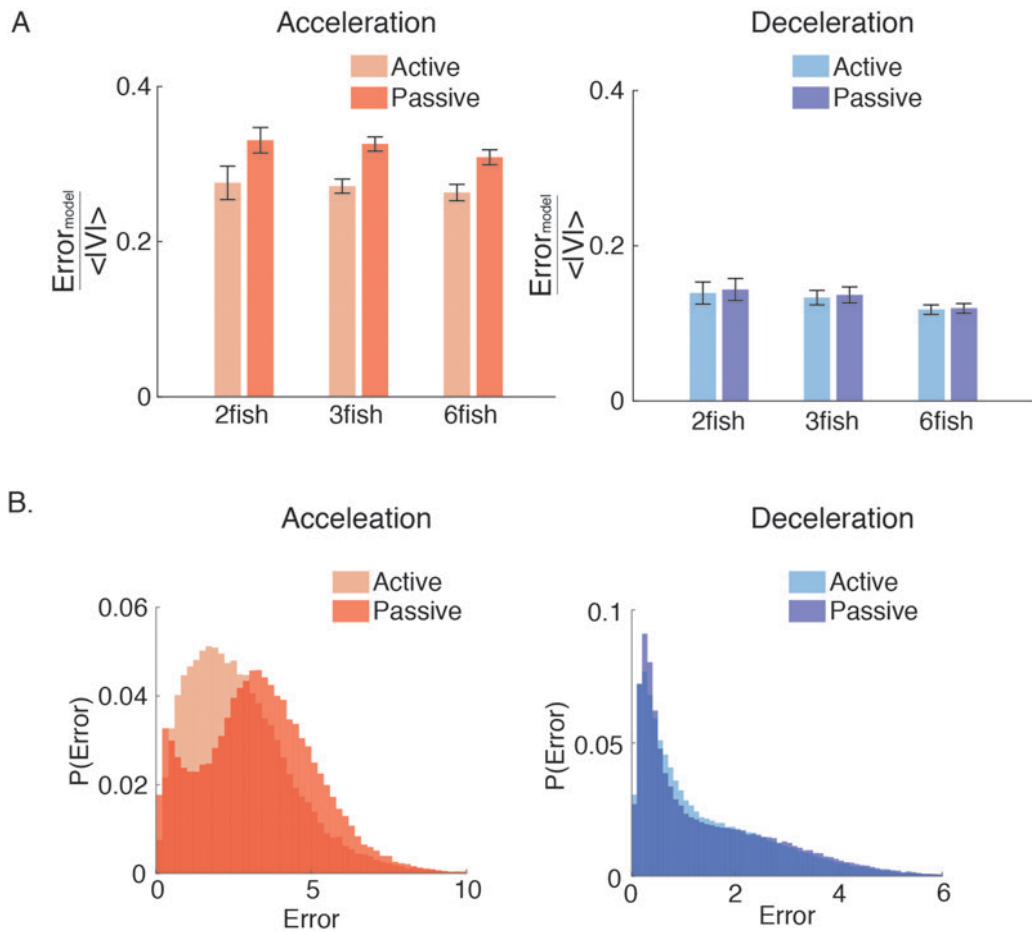

**Figure S2: Error analysis confirms that fish show little social responses during decelerations. A.** Comparison of the normalized error in prediction $|\vec{v}_{model} - \vec{v}_{real}|/\langle|\vec{v}_{real}|\rangle$ during acceleration epochs (left) and deceleration epochs (right), using an active model -- a RF model learned separately on each kinematic state (Eq. (3) main text) and a passive model (Eq. 1 main text). There is no advantage in learning a separate RF-model to predict active movements in the deceleration epochs. **B.** Example distributions of the errors $|\vec{v}_{model} - \vec{v}_{real}|$ in prediction during acceleration epochs (left) and deceleration epochs (right) for the two models as in A, again, error distributions in the deceleration epochs are similar for active or passive computation in these parts.

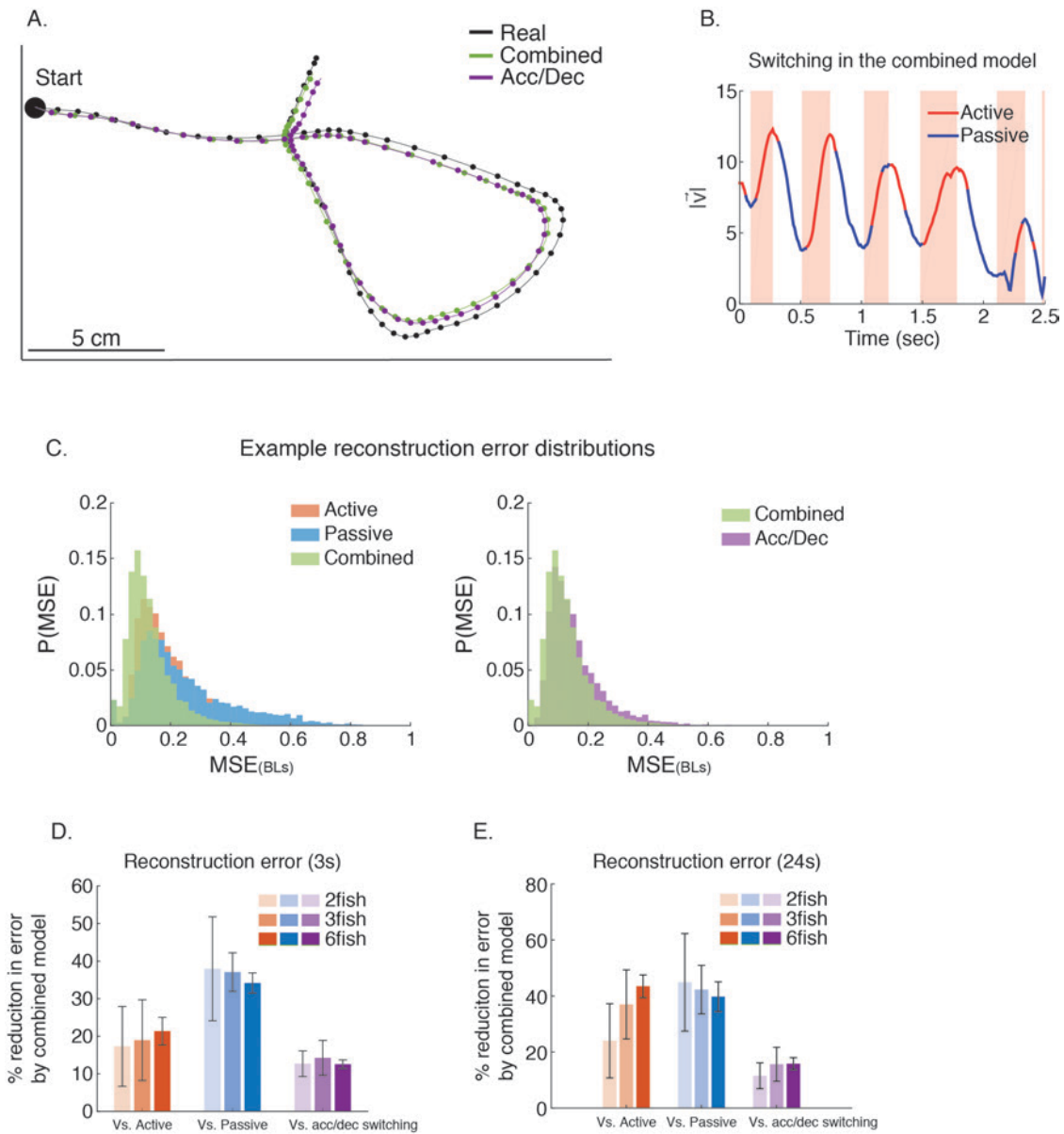

**Figure S3: Switching in combined model is comparable to the kinematic state if the fish. A.** The trajectory segment from Fig. 3C (black) and the predicted trajectory using the optimal combination of the two models (green), overlaid with the predicted trajectory based on switching between models that is done according to the acceleration or deceleration state of the fish (purple). **B.** An example of the optimal switching between models in the reconstruction shown in Fig. 3C and panel A (green line), and their high correspondence to the acceleration/deceleration state of the fish (pink background represents accelerations. **C.** Left: example distributions of the reconstruction error: $Error = \sum |\vec{c}_{model} - \vec{c}_{real}|$ using the combined model (green) compared to the active (red) and passive (blue) models. Right: comparison of the combined model (green) to switching according to the acceleration/deceleration state of the fish (purple). **D.** Average reduction in the reconstruction error of the combined model compared to the active model alone (red), the passive model alone (blue), and to the prediction obtained by switching according to the acceleration/deceleration state of the fish (purple). Error bars represent STD. **E.** Similar to D only for reconstruction of longer trajectory segments (24 sec), giving highly similar results.

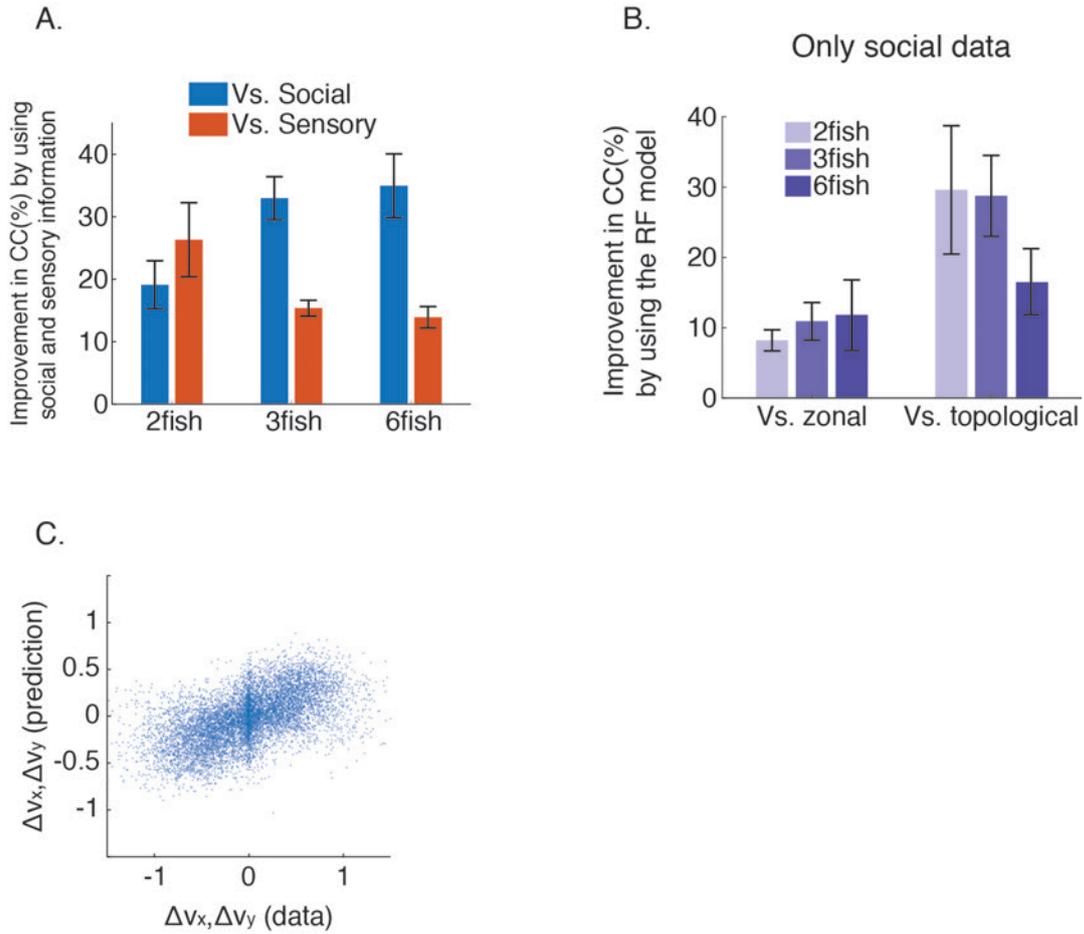

**Figure S4: Active movement changes are accurately predicted by the RF model using both social and sensory information. A.** Improvement in prediction accuracy (correlation between data and model prediction) of the RF model using both social and sensory information, compared to the social information alone (blue) and sensory information alone (red), for all group sizes (N = 6,7,7, error bars represent SEM). **B.** Improvement in prediction accuracy by using the RF-model compared to the zonal and topological models when using only sensory information for prediction for all group sizes (N=6,7,7 error bars represent SEM). **C.** Predicted velocity components $\Delta v_x, \Delta v_y$ of $\Delta \vec{v}$ plotted against their measured values. The linearity and homogeneity of variance across real values suggest we should expected a limited benefit from adding non-linarites to our model.

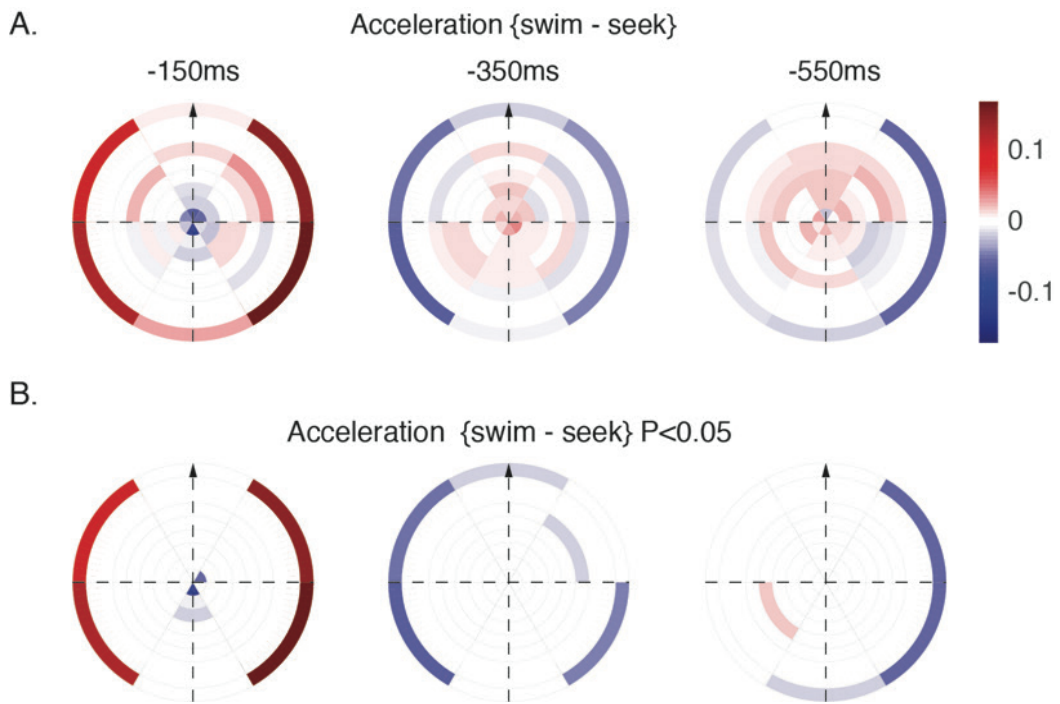

**Figure S5: Receptive field maps show distinct behavioral state dependencies. A.** Average difference maps of the receptive field for groups of 3 fish during acceleration, when performing free swimming and food seeking (see above). Bins with weights that were larger in the free-swimming state are colored red and opposite bins are in blue. **B.** Same as A, but showing only bins with differences that pass a significance test (t-test for matched samples, P<0.05).

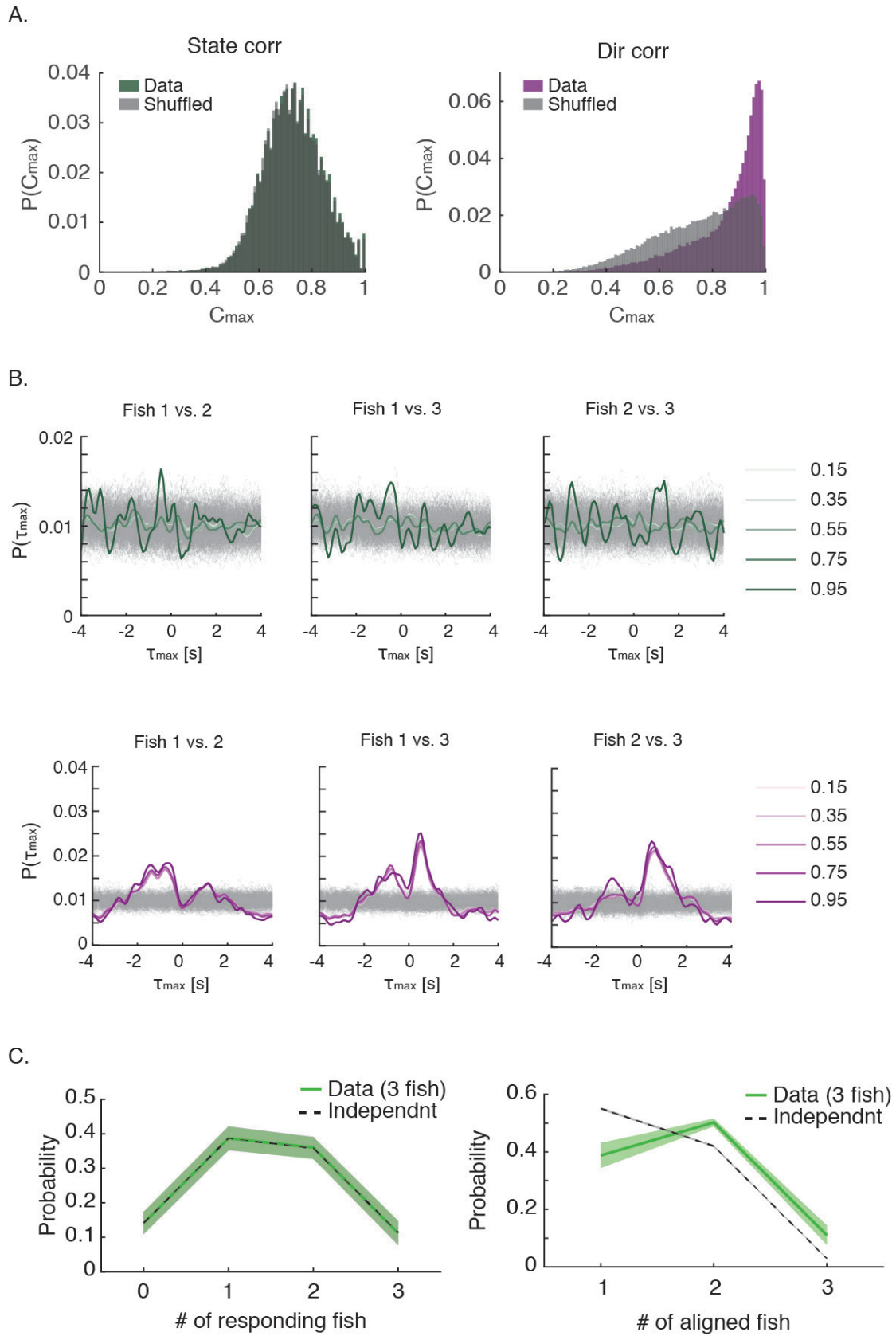

**Figure S6: Asynchronous switching between information processing modes among individual fish in a group and synchronous heading directions of group members. A.** Distributions of maximal correlation values for a group of 3 fish (from Fig 6), for the state correlation analysis (left) and for the directional correlation analysis

(right), and the correlations obtained using shuffled data in gray (see text above). **B.** Distributions of delay time ($\tau_{max}$) of the maximum correlations between pairs of fish in a group of 3 (colored lines) and shuffled controls (light gray). Different colors represent different correlation thresholds, where only $\tau_{max}$ values corresponding to maximal correlations above these thresholds are used to construct the distributions. For both the state correlation (top – green lines) and for directional correlation (bottom – purple lines), using different thresholds does not change the structure of the result presented in Fig 6. For comparison, shuffled controls are plotted using correlation threshold of 0.95. C. Group state correlation for groups of 3 fish, similar to the analysis presented in Fig. 6E-F for 6 fish.

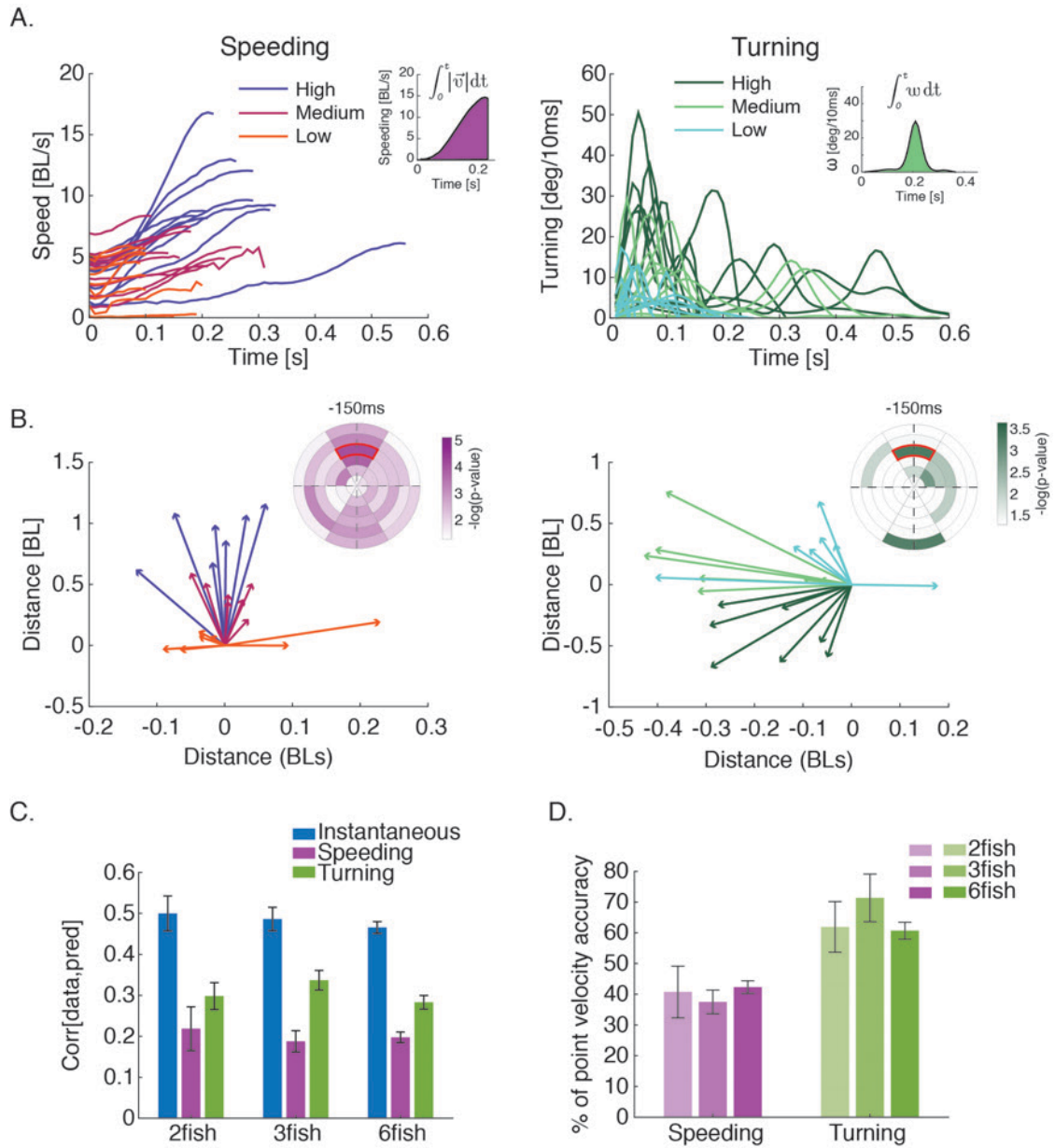

**Figure S7: Prediction of complete speeding and turning profiles using receptive field models. A.** Examples of speeding (left) and turning profiles (angular velocity - right), colored by the strength of the response ('low', 'medium', and 'high'), which were determined by discretizing the integrals of responses (insets) into even-sized bins. **B.** Inset: RF maps showing bins with significantly different neighboring fish behavior before each of the binned responses depicted in A (see methods above). Main: Average velocity vectors of the fish in the bin marked in red in the inset, for different speeding strengths (left) and turning strengths (right); strength is denoted by color as in A. **C.** Comparison of the accuracy of prediction for the instantaneous model when predicting $\Delta \vec{v}_i(t)$ (blue), and the accuracy of predicting the complete turning - $\int_0^t \omega dt$ and speeding - $\int_0^t |\vec{v}| dt$ profiles (purple and green respectively). Values are average correlation coefficient (N=6,7,7) and error bars represent SEM. **D.** Prediction accuracy

of complete turning and complete speeding responses using social and sensory information preceding these events, represented as percentage of the accuracy of predicting point accelerations ($\Delta \vec{v}_i(t)$), using similar data. Indeed, the ability to predict properties of entire acceleration epoch is significantly lower.

**References:**

1. Pérez-Escudero A, Vicente-Page J, Hinz RC, Arganda S, de Polavieja GG (2014) idTracker: tracking individuals in a group by automatic identification of unmarked animals. Nat Methods 11(7):743–748.
2. Savitzky A, Golay MJE (1964) Smoothing and Differentiation of Data by Simplified Least Squares Procedures. Anal Chem 36(8):1627–1639.
3. Tibshirani R (1994) Regression Shrinkage and Selection Via the Lasso. J R Stat Soc Ser B 58:267--288.
4. Huth A, Wissel C (1992) The simulation of the movement of fish schools. J Theor Biol 156(3):365–385.
5. Huth A, Wissel C (1994) The simulation of fish schools in comparison with experimental data. Ecol Model 75–76(0):135–146.
6. Couzin ID, Krause J, James R, Ruxton GD, Franks NR (2002) Collective memory and spatial sorting in animal groups. J Theor Biol 218(1):1–11.
7. Ballerini M, et al. (2008) Interaction ruling animal collective behavior depends on topological rather than metric distance: Evidence from a field study. Proc Natl Acad Sci 105(4):1232–1237.
8. Ákos Z, Beck R, Nagy M, Vicsek T, Kubinyi E (2014) Leadership and Path Characteristics during Walks Are Linked to Dominance Order and Individual Traits in Dogs. PLOS Comput Biol 10(1):e1003446.
9. Nagy M, Ákos Z, Biro D, Vicsek T (2010) Hierarchical group dynamics in pigeon flocks. Nature 464(7290):890–893.